# Comment on
# "Attosecond electron microscopy and diffraction"

P. Baum[1] and C. Ropers[2]

[1] *Universität Konstanz, Fachbereich Physik, 78464 Konstanz, Germany*
*Email: peter.baum@uni-konstanz.de*

[2] *Department of Ultrafast Dynamics, Max Planck Institute for Multidisciplinary Sciences,*
*37077 Göttingen, Germany*
*Email: claus.ropers@mpinat.mpg.de*

(Date: November 21, 2024)

**Abstract:** A recent paper by Hui *et al.* (Ref. [1], Sci. Adv. 10, eadp5805 (2024)) claims the demonstration of "Attosecond electron microscopy and diffraction" with laser-gated electron pulses. In this comment, we point out a series of physical and technical inconsistencies of the experiment and results. Beyond failing to show any microscopy, the reported concept does not produce properly gated electrons. Furthermore, the noise and signal levels of the presented data are statistically and quantitatively incompatible with the work's interpretation and attosecond dynamics in graphite. These inconsistencies render the claims and conclusions of Ref. [1] unsubstantiated and suggest that the data primarily show an interferometric artifact.

In Ref. [1], the authors generate femtosecond electron pulses in an ultrafast transmission electron microscope [2,3] and claim that these pulses are gated to 625-as duration by few-cycle laser pulses with time-varying elliptical polarization. Electron diffraction intensities from few-layer graphite are reported to change as a function of the delay between two collinear optical pulses on the specimen. It is claimed that these measurements demonstrate the generation of isolated attosecond electron pulses and the diffractive mapping of attosecond electron dynamics in graphite. In the following, we express and detail six fundamental concerns.

**1. Lack of electron gating:** Gating or selecting a short electron pulse out of a longer one with help of laser light requires two essential steps. First, one needs to imprint a time-dependent change onto only such electrons that have interacted with the laser light ('modulation'). Second, these electrons need to be filtered out of the remaining beam ('filtering'). This selection either requires a change in beam direction [4-6] or a change in energy [7-9] in combination with subsequent energy-resolved detection [10,11]. Modulation without filtering is not gating.

In the diffraction experiments of Ref. [1], laser-modulated electrons are not filtered out of the remaining beam. Neither an electron energy analyzer [10,11] nor a sideways deflection [4-6] nor any other beam property is applied in the experiment. Modulated electrons and unfiltered electrons therefore remain in the same beam, and both are diffracted at the graphite lattice. Reference [1] provides no reasoning for why a separation or post-selection of modulated electrons should not be necessary, while the data are discussed as if gating was achieved.

**2. Incorrect modulation principle:** Reference [1] claims that polarization control of a few-cycle laser pulse creates a single isolated attosecond electron pulse. The laser pulses have a duration of 2-3 cycles (see Fig. 3A in Ref. [1]), and phase plates let the polarization vary from elliptical to linear and back across the duration of the pulse. The interaction of these rotational laser fields with an aluminum mesh is then claimed to provide a modulation that leads to a single isolated attosecond electron pulse.

In gas-phase attosecond laser science without electron beams, high-harmonic generation with polarization-controlled few-cycle laser fields is indeed known to enable isolated attosecond photon pulses at energies of tens of eV, in the extreme ultraviolet [12,13]. The reason why polarization gating is effective in this context is the strong nonlinearity and polarization sensitivity of the high-harmonic generation process in atoms and molecules. Even a very small sideways deviation of the recolliding electron strongly weakens the spatial overlap with the remaining hole, suppressing high-harmonic generation efficiency in a strongly nonlinear way. Thus, a transient variation from elliptical to linear and back along a few-cycle pulse can be used to limit emission to a single half cycle of the driving pulse [12,13].

In contrast, the electrons in a high-energy free-space beam react to the electric field of a laser pulse in a linear way [8,9,14,15]. For swift electrons without binding to an atomic potential, the addition and removal of photon energy by laser light at a thin sample is void of recoil effects [15]. Therefore, the fundamental coupling parameter $g$ of electron-light interaction is linear in the longitudinal field component (along the electron trajectory) [7-9,16]. In the experiments of Ref. [1], a laser beam intersects a free-electron beam at an aluminum mesh, resulting in a projection of the incident elliptical field onto only one longitudinal polarization state [17]. Therefore, any kind of incoming polarization will contribute linearly to the longitudinal field component and modulate the electrons in a field-linear way. Irrespective of the details of the polarization state or the type of scattering structure, the resulting longitudinal electric field and electron-beam modulation will always contain multiple optical cycles at the same optical bandwidth as the incident laser pulse. Therefore, the reported polarization control [1] is of no advantage and will not produce isolated attosecond pulses.

**3. Unrealistic signal and noise levels:** We next show that the data presented cannot stem from sub-cycle- or even few-cycle-modulated electrons. If, for the sake of argument, the authors truly modulated 625-as electron pulses within of a sub-picosecond electron pulse, as they claim, these electrons would constitute only a fraction $P_{mod} \approx \frac{625 \text{ as}}{600 \text{ fs}} \approx 10^{-3}$ of the total beam current, a value acknowledged by the authors on page 3. In light of this figure, the reported ~5% oscillation of diffraction intensity in Fig. 5 and Fig. S3 (see description of data analysis in Supplementary Materials, page 1, and labels of Fig. S3) is clearly incompatible with sparse attosecond-modulated electrons. Moreover, irrespective of possible errors in labelling, realistic changes of diffraction intensity confined to the small temporal fraction $P_{mod}$ of the beam cannot possibly be identified using the experimental parameters of Ref. [1], because the measured Bragg spot intensity will be subject to Poisson noise of the entire beam. The estimated diffraction efficiency of six-layer graphite is $P_{Bragg} \approx 10\%$ into all six first-order spots. According to Ref. [18], the momentary change in electron diffraction due to laser-driven electron dynamics is $P_{dyn} \approx 2\%$ in an optical peak field strength of 5 V/nm. This change $P_{dyn}$ only occurs during sub-cycle times, within attoseconds. The expected change $n$ of the measured Bragg spot intensity is therefore $n = P_{dyn}P_{mod} \approx 2 \times 10^{-5}$.

The data in Figures 5 and 3S display a signal-to-noise ratio of at least $R_{SNR} \approx 5$, which would imply a sensitivity to intensity changes of the order of $n = P_{dyn}P_{mod}/R_{SNR} \approx 4 \times 10^{-6}$. Such a sensitivity is unattainable in the experiment within a realistic measurement time. The average emission current of femtosecond electron pulses is stated as $I_0 = 10^6$ electrons/s. At a total integration time $T$ per data point, the accumulated number of background electrons is $N_0 = P_{Bragg}I_0T$. The relative shot noise of this accumulated data is $1/\sqrt{N_0}$ under ideal detection efficiency. Shot noise cannot be compensated by any smoothing or referencing technique. Reducing this noise to a level lower than $n \approx 4 \times 10^{-6}$ (see above) requires a minimum integration time $T$ per data point of $T > 1/(n^2 P_{Bragg}I_0) \approx 170$ hours. Hence, the minimum necessary continuous measurement time, under ideal conditions and without technical noise, exceeds 7 days per individual data point. We arrive at about 580 days or 1.6 years for the curve shown in Fig. 5, plus the same amount again for each of the fluence-varied curves in Fig. 3S. Such measurement times are much longer than realistic in a femtosecond electron microscopy experiment. The reported signals therefore cannot stem from attosecond-modulated electrons or from attosecond electron dynamics in graphite.

**4. Optical interference artifacts:** The 'gating' beam and the pump beam in the criticized experiment are almost collinear, and the graphite sample is closely attached to the aluminum mesh. A discussion of such an experiment on the basis of one gating pulse and one pump pulse is therefore not appropriate. Both the gating pulse and the pump pulse scatter at the mesh and the graphite sample in the same way. The local electromagnetic field at the mesh and the sample is therefore a linear superposition of both fields. The total average intensity resulting from this interference depends on the relative optical delay. The thermal load [19] or the light-induced electrostatic charging of the mesh [20] depend on the optical delay. If these or other effects cause Bragg spot changes, for example by a transient or stationary Debye-Waller effect, then the mesh plus graphite sample simply act as a sensor of total cycle-averaged optical intensity, as it is measured in a Michelson or Mach-Zehnder interferometer trace.

The reported large signal amplitudes of 5% and above (supplementary material and original arXiv version of Ref. [1]) are not only incompatible with attosecond effects (see above) but also strongly suggest an interferometric artifact and a modulation of the entire, ungated electron beam. Notably, Fig. 2 of the arXiv version of Ref. [1] reports an increasing modulation amplitude for larger scattering vectors (i.e., higher-order peaks), from about 4% (first-order peaks) to approximately 10% (third-order peaks), which points to a Debye-Waller effect. Figure 5 in the journal version of Ref. [1] shows the same data curves and axis label, however, with percentages replaced by 'arb. unit'.

None of the authors' cross checks (see 'confirmation results') are capable of ruling out a plain interference effect: Removing the gating medium changes and likely reduces the near-field scattering and the cycle-averaged response of the specimen, such as absorption. Blocking the gating pulse or the pump pulse obviously prevents the appearance of any oscillations in the data trace, because there is no longer a second pulse.

**5. Microscopy:** The authors claim that their experiment can be called attosecond electron microscopy ('attomicroscopy'). However, microscopy refers to the production of magnified images of a material in spatial coordinates, a result that is not achieved.

**6. Lack of due diligence:** A large range of standard parameters are not mentioned or discussed in Ref. [1]. For example, the authors do not provide the optical focus diameter, the diameter of the electron beam, its divergence and spatial coherence, the electron pulse duration, the electron detection efficiency, or the total integration time. Neither a measurement of gated electron pulses nor raw data is provided, and a quantitative evaluation is hindered by the use of arbitrary units for the measured and simulated diffraction effects. The authors do not change the temporal overlap with the electron pulse to check for conventional pump-probe dynamics, although such measurements are standard in ultrafast spectroscopy and would directly reveal the above-mentioned inconsistencies.

In summary, the authors of Ref. [1] did not perform microscopy, did not produce gating or isolated attosecond pulses and did not measure sub-cycle electron dynamics in graphite, but carried out experiments with interferometric, technical artifacts. We note that each of these issues is also present in a recent arXiv manuscript by the same authors [21].